\documentclass[sigconf]{acmart}

\usepackage{adjustbox}

\AtBeginDocument{%
  \providecommand\BibTeX{{%
    \normalfont B\kern-0.5em{\scshape i\kern-0.25em b}\kern-0.8em\TeX}}}

\copyrightyear{2021}
\acmYear{2021}
\setcopyright{acmcopyright}\acmConference[SAC '21]{The 36th ACM/SIGAPP Symposium on Applied Computing}{March 22--26, 2021}{Virtual Event, Republic of Korea}
\acmBooktitle{The 36th ACM/SIGAPP Symposium on Applied Computing (SAC '21), March 22--26, 2021, Virtual Event, Republic of Korea}
\acmPrice{15.00}
\acmDOI{10.1145/3412841.3442011}
\acmISBN{978-1-4503-8104-8/21/03}

\usepackage{systeme}
\usepackage{xspace}
\usepackage{algorithm}
\usepackage[noend]{algpseudocode}
\usepackage{multirow}
\usepackage{comment}
\usepackage{subcaption}
\usepackage{booktabs}
\usepackage[normalem]{ulem}
\begin{document}

% Algos
\newcommand{\algoname}[1]{\textsc{{#1}}\xspace}
\newcommand{\puresvd}{\algoname{PureSVD}}
\newcommand{\itemknn}{\algoname{ItemKNN}}
\newcommand{\userknn}{\algoname{UserKNN}}
\newcommand{\slim}{\algoname{SLIM}}

\newcommand{\bprmf}{\algoname{BPR-MF}}
\newcommand{\bprnnmf}{\algoname{BPR-NNMF}}

\newcommand{\funkmf}{\algoname{Funk-MF}}
\newcommand{\funknnmf}{\algoname{Funk-NNMF}}

\newcommand{\pmf}{\algoname{P-MF}}
\newcommand{\pnnmf}{\algoname{P-NNMF}}
\renewcommand{\vec}[1]{\mathbf{#1}}

% Datasets
\newcommand{\datasetname}[1]{{#1}\xspace}
\newcommand{\lastfm}{\datasetname{LFM}}
\newcommand{\movielensom}{\datasetname{M1M}}
\newcommand{\bookcrossing}{\datasetname{BCR}}
\newcommand{\pinterest}{\datasetname{PIN}}
\newcommand{\epinions}{\datasetname{EPI}}
\newcommand{\citeulike}{\datasetname{CUL}}

%% annotations
\newcommand{\gabbo}[1]{\textcolor{cyan}{\textbf{[Giovanni] #1}}}
\newcommand{\cb}[1]{\textcolor{red}{\textbf{[Cesare] #1}}}
\newcommand{\edo}[1]{\textcolor{magenta}{\textbf{[Edoardo] #1}}}

%%
%% The "title" command has an optional parameter,
%% allowing the author to define a "short title" to be used in page headers.
\title{Improving repeatably in Matrix Factorization algorithms for Top-N recommendations}
\title{Improving stability in Matrix Factorization algorithms for Top-N recommendations}
\title[Improving stability in Matrix Factorization for Top-N recommendations]{Improving stability in Matrix Factorization \\ for Top-N recommendations}
\title[On the instability of embedding-based recommenders: \\the case of Matrix Factorization]{On the instability of embedding-based recommenders: the case of Matrix Factorization}
\title[On the instability of embeddings for recommender systems: \\the case of Matrix Factorization]{On the instability of embeddings for recommender systems: \\the case of Matrix Factorization}
%%
%% The "author" command and its associated commands are used to define
%% the authors and their affiliations.
%% Of note is the shared affiliation of the first two authors, and the
%% "authornote" and "authornotemark" commands
%% used to denote shared contribution to the research.

\author{Giovanni Gabbolini}
\authornote{The work was mainly carried out while the authors were affiliated with Politecnico di Milano. The two authors have contributed equally to the work.}
\affiliation{
\institution{Insight Centre for Data Analytics}
\city{University College Cork, Ireland}
}
\email{giovanni.gabbolini@insight-centre.org}

\author{Edoardo D'Amico}
\authornotemark[1]
\affiliation{
\institution{Insight Centre for Data Analytics}
\city{University College Dublin, Ireland}
}
\email{edoardo.d'amico@insight-centre.org}

\author{Cesare Bernardis}
\orcid{0000-0002-8972-0850}
\affiliation{%
  \city{Politecnico di Milano, Italy}
}
\email{cesare.bernardis@polimi.it}

\author{Paolo Cremonesi}
\orcid{0000-0002-1253-8081}
\affiliation{%
%   \institution{DEIB}
  \city{Politecnico di Milano, Italy}
}
\email{paolo.cremonesi@polimi.it}

% \affiliation{%
%   \institution{DEIB, Politecnico di Milano}
%   \city{Milan}
%   \state{Italy}
% }

%%
%% By default, the full list of authors will be used in the page
%% headers. Often, this list is too long, and will overlap
%% other information printed in the page headers. This command allows
%% the author to define a more concise list
%% of authors' names for this purpose.
\renewcommand{\shortauthors}{G. Gabbolini, E. D'Amico et al.}

%%  custom commands
\newcommand{\citequote}[1]{``#1''}
\newcommand{\scalar}{\text{\large{\textbf{\textperiodcentered}}}}
\newcommand{\nn}[1]{\textrm{NN}_{#1}}

%%
%% The abstract is a short summary of the work to be presented in the
%% article.
\begin{abstract}

Most state-of-the-art top-N collaborative recommender systems work by learning embeddings to jointly represent users and items. Learned embeddings are considered to be effective to solve a variety of tasks. Among others, providing and explaining recommendations.
In this paper we question the reliability of the embeddings learned by Matrix Factorization (MF).
We empirically demonstrate that, by simply changing the initial values assigned to the latent factors, the same MF method generates very different embeddings of items and users, and we highlight that this effect is stronger for less popular items.
To overcome these drawbacks, we present a generalization of MF, called Nearest Neighbors Matrix Factorization (NNMF).
%The new method learns the latent representation of each user and item as a linear combination of the representations of its neighbors.
%This strategy has the effect to propagate the information of items and users to their neighbors, speeding up the training procedure and extending the amount of information that supports recommendations and representations.
The new method propagates the information about items and users to their neighbors, speeding up the training procedure and extending the amount of information that supports recommendations and representations.
We describe the NNMF variants of three common MF approaches, and with extensive experiments on five different datasets we show that they strongly mitigate the instability issues of the original MF versions and they improve 
the accuracy of recommendations on the long-tail.

\end{abstract}



%%
%% Keywords. The author(s) should pick words that accurately describe
%% the work being presented. Separate the keywords with commas.

%% A "teaser" image appears between the author and affiliation
%% information and the body of the document, and typically spans the
%% page.

%%
%% This command processes the author and affiliation and title
%% information and builds the first part of the formatted document.
\maketitle

\section{Introduction}
\label{sec:introduction}
The main goal of a wide variety of top-N collaborative recommender systems is to learn embeddings to jointly represent users and items \cite{puresvd, he2017neural, xue2017deep}. 
Embeddings are derived in such a way that similar items (or similar users) have similar embeddings \cite{liang2016factorization}.
Embeddings can be learnt with either Matrix Factorization (MF) or deep learning (DL) techniques \cite{koren2009matrix, he2017neural}.
For both families of techniques, the learning procedure requires to sample the interaction space during the training (e.g., stochastic-gradient-descent \cite{ruder2016overview}, skip-grams with negative sampling \cite{mikolov2013linguistic}).

Embedding-based models are characterized by various sources of randomness in their training, such as the initial values of the embeddings and the sampled interactions.
Different random seeds or random generators might lead to different results in terms of embeddings learnt and, consequently, items recommended.
The magnitude of these differences determines whether an algorithm can be considered \textit{stable} or not.
%We consider an algorithm to be \textit{stable} if these differences are within a tolerance range.
If these differences are too large (i.e, if the algorithm is \textit{not stable}), we incur in severe issues that could mine the reliability and the credibility of a recommender.
First, we cannot consider its recommendations to be reliable anymore, as the same algorithm executed on the same dataset could make totally different predictions for the same user-item pair \cite{madhyastha2019model}.
Second, the instability of users' and items' representations may have strong implications in the explainability of the recommendations.
Interpreting the predictions of embedding models is a difficult task, because there is not a clear and direct relationship between the representations learnt by the model and the attributes or the interactions of users and items.
Most solutions provide explanations identifying similarities between the embeddings obtained by the model \cite{tintarev2007survey, abdollahi2017using}, but if the representations of users and items in the embedding space are not stable, the quality of such explanations might be compromised.
Third, our definition of stability is strictly connected to that of \emph{repeatability} as defined, for instance, in the SIGIR Initiative to implement the ACM Artifact Review and Badging guidelines\footnote{\url{acm.org/publications/policies/artifact-review-and-badging-current}}.
According to this definition, an unstable algorithm is not repeatable.

Alongside these examples, the works that leverage the embeddings of a model to various extents are not concerned with the potential consequences of this instability.
Indeed, our notion of stability of a recommender system is different from what has been commonly addressed in the literature. 
Some works focus on the stability of the model when you introduce noisy or malicious perturbations in the dataset \cite{said2012users, mobasher2007attacks}, or when it evolves naturally with more interactions \cite{adomavicius2014improving, adomavicius2012stability}.
In this work we focus on the stability of a recommender system when the same model is trained on the same dataset in exactly the same experimental conditions but with a different random sequence (e.g., due to a different random seed).

The variance in both the internal representation of the model (embeddings) and the output of the model (estimated relevance) can be reduced if enough samples are collected for each user and item to be modeled.
Unfortunately, most datasets exhibit strong popularity biases \cite{anderson2006long}. 
Because of these biases, unpopular items (i.e., items with very few interactions) and short-profile users (i.e., users with very few interactions) will not collect enough samples during their training to smooth out the noise introduced by the randomness.
Embeddings learnt on a small sample size will be biased and unstable \cite{skurichina1998bagging}. 
%As such, we expect the embeddings of unpopular items and short-profile users to vary greatly across different runs of the same experiment.

Different techniques exist to improve the generalization capabilities of a model by leveraging or controlling the randomness of the training procedure.
Bagging \cite{skurichina1998bagging} is an ensemble method that trains different models from boostrap replica of the same dataset and average their predictions.
However, it requires to retrain the model several times and is designed to improve the generalization capabilities and not to stabilize the model.
%Embeddings learnt on very small sample sizes have bad expected quality and large instability. 
Other techniques, such as Stochastic Weight Averaging (SWA), produce an ensemble by averaging the weights of the same model at different epochs of the training process \cite{madhyastha2019model,izmailov2018averaging}.
In case of linear models, such as with Matrix Factorization, this is equivalent to averaging the embedding vectors during the gradient descent optimization. 
%The esembling is performed across different snaphshots of the same embedding during the training.
%ensembles in the model space
%combine several models and then use models’ predictions to produce the final prediction.
% ensembling in the weights space. 
%This method produces an ensemble by combining weights of the same network at different stages of training 
All these techniques are model agnostic and do not take into account the neighborhood properties of user and item embeddings: similar items (users) have similar embeddings.

In this paper, we focus on the stability of a widely employed family of embedding-based models, that is Matrix Factorization, as a function of random-seed.
We are especially interested in investigating if MF based models under different initializations of latent factors generated by distinct random seeds, produce similar predictions and similar embeddings.
We propose a new framework called Nearest Neighbors Matrix Factorization (NNMF), a generalization of classic Matrix Factorization that merges MF with nearest neighbors (NN) in order to alleviate the effects brought by the scarcity of interactions for unpopular items.
While in classic MF the latent representations of items or users are treated independently, in NNMF we force each embedding to be a linear combination of the embeddings of a set of their most similar neighbors.
With this approach we map the neighborhood relationships among items or users from the original interaction space to the new latent space.
This mapping has the effect to propagate the updates applied to an item or a user also to their respective neighbors.
%Thanks to this property, less popular (i.e., less sampled) items or users get updated more often, e.g., whenever a neighbor item or user is updated.

We provide an extensive set of experiments, testing NNMF with three MF approaches (\bprmf, \funkmf and \pmf) over five different datasets. 
The results show that: 
\begin{itemize}
\item 
all the three MF methods suffer from the instability problem:
on average, recommended items change by more than 50\%; %by simply retraining the same model with a different random seed;
%and that are not always more accurate on the long-tail with respect to common collaborative baselines.
\item 
the three NNMF variants greatly improve stability: on average, recommended items change by less than 25\%;
\item 
the NNMF variants have better accuracy on the long tail with respect to the original MF methods, as well as other baselines, in almost all measures and datasets;
\item
the improved stability and the information propagation of NNMF allow to reach convergence in a fraction of the number of epochs required by MF. 
\end{itemize}

The rest of the paper is organized as follows.
In Section \ref{sec:related-works} we list a set of works related to the argument we treat in this paper.
In Section \ref{sec:models} we introduce the issues of classic Matrix Factorization techniques and we present our new framework called Nearest Neighbor Matrix Factorization. We also describe some practical implementations over three well known MF algorithms.
In Section \ref{sec:experiments} we discuss the results obtained by an extensive set of experiments over a variety of different datasets, testing the new models under different aspects.
Finally, in Section \ref{sec:conclusions}, we provide some concluding remarks.

\section{Related Work}
\label{sec:related-works}

There exist different definitions of stability of a recommender system in the literature, and different ways to improve each of these definitions.
Most works define the stability of a recommender system as the "consistent agreement of predictions" made to the same user by the same algorithm, when new incoming interactions  are added to the system in complete agreement to system’s prior predictions \cite{adomavicius2012stability}. 
For instance, the work in \cite{adomavicius2014improving} adopts bagging and iterative smoothing in conjunction with different traditional recommendation algorithms to improve their consistency. 
Other works define stability as the ability of the recommender system to provide consistent recommendations when malicious perturbations are performed to the dataset \cite{adomavicius2012stability}.
The work in \cite{mobasher2007attacks} suggests hybrid collaborative and content-based filtering as the best solution to mitigate the effects of attacks on the consistency of recommendations. 
Finally, other works \cite{said2018coherence} relate the stability, or confidence, of a recommender system with the quality of a dataset, either at system level (the magic barrier described in \cite{said2018coherence}) or at user-level \cite{bernardis2019eigenvalue}.
Our notion of stability -- the consistency of both recommendations and latent representations of users and items when the same model is trained on exactly the same dataset with a different random sequence -- is different from the definitions used in the literature. 

There are also several works that try to control (or leverage) the randomness intrinsic in machine learning algorithms in order to improve the generalization capabilities of a model.
Bagging is the most widely adopted black-box method used to leverage randomness in the input data in order improve the classification accuracy of a model \cite{skurichina1998bagging}. 
Bagging builds an ensemble of models by (i) running the same training algorithm on different boostrap replica of the same dataset and (ii) by aggregating their predictions. 
Training multiple model for prediction averaging, as with bagging, is computationally expensive. 
Therefore, other works train a single model and save the model parameters (snapshots) along the optimization path. 
The predictions of the snapshot models are later combined to produce the final prediction \cite{huang2017snapshot}.
Differently from bagging and snapshots, that build ensambles in the model space, other works build ensembles in the weights space. 
For instance, the works in \cite{madhyastha2019model} and \cite{izmailov2018averaging} use two variants of the same technique, Stochastic Weight Averaging (SWA), to compute a running average of the model weights during the last epochs of the training process.

Note also that even though the idea to merge MF and NN is not new \cite{nova2014nnmf, koren2008nnmf}, this is the first work, to the best of our knowledge, that addresses this particular stability issue and proposes a generic framework for MF that alleviates its drawbacks.

\section{Models} 

\label{sec:models}

\label{subsec:preliminaries}
In the following Sections, we denote with $\mathcal{U}$ and $\mathcal{I}$ the sets of users and items and with $|\mathcal{U}|$ and $|\mathcal{I}|$ their cardinalities.
Lower case letters $u, v$ will be used to refer to users, while $i, j, k$ will refer to items.
The user rating matrix (URM) is indicated by the uppercase bold letter $\vec{R}$ and each cell $r_{ui}$ contains the value of the preference, either explicit or implicit, if expressed, 0 otherwise.
Lower case, bold letters indicate vectors in column format, unless differently specified.
In particular, $\vec{r_{u}}$ indicates the user profile, intended as the $u$-th row of matrix $\vec{R}$.
%, while  $\vec{r_{i}}$ indicates the item profile, intended as the $i$-th column of $\vec{R}$.
The set of $(u,i)$ couples for which $r_{ui}$ is known is denoted with $\kappa$.

In its basic form, MF consists in representing users and items in a latent factor space of dimension $f$. In particular, users and items are represented respectively in matrices $\vec{P} \in \mathbb{R}^{|\mathcal{U}| \times f}$ and $\vec{Q} \in \mathbb{R}^{|\mathcal{I}| \times f}$. A row $\vec{p_u}$ of $\vec{P}$ and $\vec{q_i}$ of $\vec{Q}$ handle, respectively, the representation of a user and an item on the same latent factor space. The dot product
\begin{equation} \label{eq:basic_mf}
    \overline{r}_{ui} = \vec{p_u} \scalar \vec{q_i^T}
\end{equation}
measures how much user $u$ and item $i$ are aligned in the new latent space and therefore an estimate of the rating $\overline{r}_{ui}$ is provided. Consequently, the user-rating matrix can be estimated as:
\begin{equation}\label{eq:basic_mf_matrix_form}
\overline{\vec{R}} = \vec{P} \vec{Q^T}
\end{equation}

\subsection{Matrix Factorization drawbacks}
\label{subsec:mf-problems}
Over the years, researchers proposed various MF algorithms, varying how matrices $\vec{P}$ and $\vec{Q}$ are learned \cite{funk, BPR, probmf} in order to improve the quality of recommendations under different aspects.
Most of them learn the parameters of the model optimizing an objective function through stochastic gradient descent, iterating over the available data.
The training phases of such algorithms share a common schema, which is partially altered from case to case.
Firstly the two latent factor matrices $\vec{P}$ and $\vec{Q}$ are initialized with random values, then an iterative learning procedure begins.
With each iteration, one or more interactions are selected among the available ones, and the objective function is computed alongside its respective gradient. 
Finally, the latent factors of the sampled items and users are updated accordingly.

%Notice that in this type of learning procedure, the impact of missing information, which is the vast majority, is not taken into account.
%In some cases, like the Bayesian Personalized Ranking optimization procedure\cite{BPR}, the missing as negative assumption is made \footnote{Missing as negative means that the absence of a preference for a user towards an item is transformed into a negative preference}, and a part of missing interactions is considered into the learning phase as negative preferences.
%However, only a small part of the whole missing information is considered and the missing as negative assumption, even if reasonable, is quite strong and might be wrong in some cases.

The interactions are strongly biased towards popular items and long profiles.
As such, the latent representations of users and items are updated based on this biased distribution: factors belonging to popular items (users) are updated far more often than niche ones.
This means that the random initialization of the latent representations values for  unpopular items and short-profile users has a strong impact on their final representations at convergence. We refer to this issue as \textit{instability of representations}.

This problem directly affects also the recommendation lists generated by the algorithms:
if the representation of a user is unstable, also the closest items in the latent space are unstable, leading to the generation of different recommendation lists for the same user, based on different random initial conditions. We refer to this issue as \textit{instability of recommendations}.
%In Section \ref{sec:experiments} we empirically show that very often the same model, trained on the same data, leads to different representations for items and users and, consequently, to different recommendation lists generated.
\begin{table}[t]
\small
\centering
\caption{Stability of Top-10 recommendation lists generated by three MF techniques expressed with Jaccard index. Higher values indicate a better stability. Datasets names abbreviations are reported in Section \ref{sec:datasets}.}
\begin{tabular}{c|c|c|c|c|c}
\toprule   
\multirow{1}{*}{\textbf{Algorithm}} & \multicolumn{1}{c|}{\textbf{\lastfm}} & \multicolumn{1}{c|}{\textbf{\movielensom}} & \multicolumn{1}{c|}{\textbf{\bookcrossing}} & \multicolumn{1}{c|}{\textbf{\pinterest}} & \multicolumn{1}{c}{\textbf{\citeulike}} \\ \midrule 
 \bprmf & 0.70 & 0.78 & 0.28 & 0.50 & 0.49 \\
 \funkmf & 0.72 & 0.60 & 0.05 & 0.28 & 0.31 \\
 \pmf & 0.62 & 0.60 & 0.23 & 0.39 & 0.37 \\
\bottomrule
\end{tabular} 
\label{table:stability_recommendations_mfonly}
\end{table}
As an example, in Table \ref{table:stability_recommendations_mfonly} we show the stability of the top-10 recommendations for three common MF techniques, expressed with the Jaccard index, that indicates how much the generated lists overlap.
The details of the experimental setup are described in Section \ref{sec:experiments}.
We compare the recommendations provided by 10 instances of the same model, all trained on the same data and with the same configuration, changing only the random seed, which, in turn, affects the initial values of the latent factors.
The results in Table \ref{table:stability_recommendations_mfonly} show that, according to the Jaccard index, the recommendation lists overlap by less than 50\% for three datasets out of five.
For these datasets, more than 50\% of the recommended items change by changing the initial random seed, i.e. by simply altering the initial values of the latent factors.
For the BookCrossing dataset the instability is even more dramatic: lists overlap by less than 30\%.
More details about the experimental procedure and a wider range of experiments are described in Section \ref{sec:experiments}.

\subsection{Nearest Neighbors Matrix Factorization}
The objective of MF is to learn users and items representations, also called embeddings, in a new latent space where users are mapped close to items they have expressed positive preference for.
MF algorithms perform this task treating users and items as independent entities, without explicitly taking into account existing relationships among users and among items.
However, these meaningful relationships should be reproduced also in the new latent space.
%Consequently, we could propagate the information we learn for a user or an item also to their respective neighbors, with a weight proportional to how users and items involved were similar in the original space.
We hence propose a new framework called \textit{Nearest Neighbors Matrix Factorization} (NNMF), which is able to let Matrix Factorization algorithms leverage knowledge about users and items relationships, under the form of similarities, during the algorithm learning procedure.
%The result are users and items embeddings which are directly coupled with each other, to which we refer to as \textit{neighborhood-aware} latent factor representations.
Given the two latent factor matrices $\vec{P}$ and $\vec{Q}$ defined in Section \ref{subsec:preliminaries}, we define the new neighborhood-aware latent representations for users $\vec{P^*}$ and items $\vec{Q^*}$ 
\begin{equation*}
    \vec{P^*} = \vec{S^U} \vec{P}
    \hspace{30pt}
    \vec{Q^*} = \vec{S^I} \vec{Q}
\end{equation*}
where $\vec{S^U}$ and $\vec{S^I}$ are a user similarity matrix and an item similarity matrix, respectively, the first with size $|\mathcal{U}| \times |\mathcal{U}|$ and the second with size $|\mathcal{I}| \times |\mathcal{I}|$.
Each element $s_{xy}$ stores the value of the similarity between entity $x$ and $y$, being them either users or items (i.e. $x,y \in \mathcal{U}$ or $x,y \in \mathcal{I}$).
We require $s_{xx}$ to be equal to 1 for any $x$.
Notice that if both similarity matrices $\vec{S^U}$ and $\vec{S^I}$ are identity matrices, NNMF collapses to classic MF.
It follows that the new NNMF framework is a generalization of MF.

An important characteristic of the new technique can be highlighted by exploding the neighborhood-aware representations
%, as shown in  \eqref{eq:na-representation-exploded-user} and \eqref{eq:na-representation-exploded-item}.
\begin{equation}
\label{eq:na-representation-exploded-user}
    \vec{p^*_{u}} = \sum_{v \in \mathcal{U}}s^U_{uv} \vec{p_v} =  \vec{p_u} + \sum_{v \in \mathcal{U} \setminus \{u\}} s^U_{uv} \vec{p_v}
\end{equation}
\begin{equation}
\label{eq:na-representation-exploded-item}
    \vec{q^*_i} = \sum_{j \in \mathcal{I}} s^I_{ij} \vec{q_j} =  \vec{q_i} + \sum_{j \in \mathcal{I} \setminus \{i\}} s^I_{ij} \vec{q_j}
\end{equation}
We can clearly distinguish the contributions of the independent factors $\vec{p_u}$ and $\vec{q_i}$ from the contributions of the neighborhoods embeddings.
The magnitude of the contribution of each neighbor is defined by the similarity with the user or the item we are considering: the more similar they are, the stronger the contribution will be.
Finally, we can modify \eqref{eq:basic_mf_matrix_form}, used to estimate the preferences and provide recommendations, with the new formulation of the latent representations, rewriting it as:
\begin{equation}\label{eq:basic_nnmf_matrix_form}
\overline{\vec{R}} = \vec{P^*} \vec{Q^{*T}} = \vec{S^U} \vec{P} \vec{Q^T} \vec{S^{I^T}}
\end{equation}
\begin{equation} \label{eq:basic_nnmf}
    \overline{r}_{ui} = \vec{p^*_u} \scalar  \vec{q^{*T}_i} = \bigg( \sum_{v \in U} s^U_{uv} \vec{p_v} \bigg) \scalar \bigg( \sum_{j \in I} s^I_{ij} \vec{q_j} \bigg)
\end{equation}
In our implementation, the similarity matrices $\vec{S^U}$ and $\vec{S^I}$ are constant matrices that can be pre-computed by using any traditional nearest-neighbor collaborative-filtering approach.
As such, NNMF can be easily applied to almost any MF algorithm.

Note that with the proposed formulation, $\vec{P}$ and $\vec{Q}$ do not directly contain users and items embeddings.
They contain, instead, vectors that form a generating set, not necessarily a basis, for the vector space where users and items representations are projected.
Embeddings for users and items are now contained in $\vec{P^*}$ and $\vec{Q^*}$, respectively.
Another important aspect to notice is how an estimated rating $\overline{r}_{ui}$ is now dependent on multiple users and items and it is not restricted to $u$ and $i$ anymore.

There are two main advantages with NNMF over traditional MF.
The first advantage is that it allows to have a larger amount of information supporting the latent representations of items and users, a particularly important aspect for users and items that have scarce data available, leading to a higher stability of recommendations and representations.
The second advantage is that the updates made to the embeddings during the learning procedure are not restricted to the user and the item associated to the sampled interaction, but are also propagated to the representations of users and items in the neighborhoods of $u$ and $i$, resulting in a faster convergence of the model.

\newif\ifinstances
%\instancestrue

\section{Experiments}
\label{sec:experiments}

\subsection{Similarity}
In the NNMF algorithm, relationships among users and items are modeled in the form of similarity matrices $\vec{S^U}$ and $\vec{S^I}$.
Even though we did not make any assumption on how these matrices are obtained, for the experimental part of this paper we assume that the similarity values are calculated using the \textit{shrinked cosine similarity} function:
\begin{equation} \label{eq:cosine}
    %s_{uv} = \frac{\sum_i^{\mathcal{I}} r_{ui} \cdot r_{vi}}{\sqrt{\sum_i^{\mathcal{I}}} r_{ui}^2 \sqrt{\sum_i^{\mathcal{I}}} r_{vi}^2 + h}
    s_{uv} = \frac{\vec{r_u} \cdot \vec{r_v}}{||\vec{r_u}|| \cdot ||\vec{r_v}|| + h_U}
    \hspace{2cm}
    %s_{ij} = \frac{\sum_u^{\mathcal{U}} r_{ui} \cdot r_{uj}}{\sqrt{\sum_u^{\mathcal{U}}} r_{ui}^2 \sqrt{\sum_u^{\mathcal{U}}} r_{uj}^2 + h}
    s_{ij} = \frac{\vec{r_i} \cdot \vec{r_j}}{||\vec{r_i}|| \cdot ||\vec{r_j}|| + h_I}
\end{equation}
where $\vec{r_u}$ and $\vec{r_v}$ are user profiles, $\vec{r_i}$ and $\vec{r_j}$ are item profiles and $h_U$ and $h_I$ are the shrink terms.
Moreover, for every item and user we kept only a small number of the nearest neighbors, since we noticed that this approach led to the best performance.

Note that the choice of the \textit{cosine} has two main advantages.
The first is that it is simple and fast to compute.
The second is that this way, the NNMF model has the same complexity of the original MF ones and also the same number of parameters to learn.
%The potential improvements introduced by more refined similarity functions or by the inclusion of the similarity values in  $\vec{S^U}$ and $\vec{S^I}$ as new parameters to learn will be subject of future studies.

\subsection{Instances}
We experimented NNMF with three well known MF algorithms: \funkmf \cite{funk}, \bprmf \cite{BPR} and \pmf \cite{probmf}.

\begin{itemize}
\item The NNMF loss function for \funkmf is
\begin{equation*}  \label{eq:loss_funknnmf}
    \sum_{(u,i)} \bigg(r_{ui} - \bigg( \sum_{v} s_{uv} \vec{p_v} \bigg) \scalar \bigg( \sum_{j} s_{ij} \vec{q_j} \bigg) \bigg) + \lambda_p \sum_{v} ||\vec{p_v}||^2 + \lambda_q \sum_{j} ||\vec{q_j}||^2
\end{equation*}
where $\lambda_p$ and $\lambda_q$ control the regularization. 

\item The maximum posterior estimator for the NNMF version of \bprmf is
\begin{equation*} \label{eq:loss_bprnnmf}
\sum_{(u,i,j)}{\ln{\sigma(\overline{x}_{uij}(\Theta))}} + \lambda_{\Theta}\left\Vert\Theta\right\Vert^2
\end{equation*}
where $\lambda_{\Theta}$ are model specific regularization parameters and $\sigma ( \cdot )$ is the logistic function.
The difference to the original MF version is in how $\overline{x}_{uij}(\Theta)$ is calculated, that for NNMF is
\begin{equation*} \label{eq:xuij_bprnnmf}
    \overline{x}_{uij}(\Theta) = %\overline{r}_{ui} - \overline{r}_{uj} = 
    \bigg( \sum_{v} s_{uv} \vec{p_v} \bigg) \scalar \bigg( \sum_{k} s_{ik} \vec{q_{k}} \bigg) -
    \bigg( \sum_{v} s_{uv} \vec{p_v} \bigg) \scalar \bigg( \sum_{k} s_{jk} \vec{q_{k}} \bigg)
\end{equation*}

\item The loss of the NNMF version of \pmf is
\begin{equation} \label{eq:loss_nnpmf}
    \sum_{u, i} \bigg( r_{ui} - \sigma \bigg( \sum_{v} s_{uv} \vec{p_v} \scalar \sum_{k} s_{ik} \vec{q_k} \bigg) \bigg) ^2 + \frac{\lambda_p}{2} \sum_{u} ||\vec{p_u}||^2 + \frac{\lambda_q}{2} \sum_{i} ||\vec{q_i}||^2 
\end{equation}
where $\lambda_p$ and $\lambda_q$ are the regularization parameters.

\end{itemize}

In the experiments we forced all the NNMF models to use at least 2 neighbors for users or items, in order to ensure that a difference exists between the MF and the NNMF instances of the same approach\footnote{If the number of neighbors for items and users is 1, NNMF is equivalent to MF}.
All the algorithms have been trained using an early stopping technique based on the accuracy performance on the validation set.
We also performed a Bayesian optimization on a validation set to find the best parameters for every approach we tested.
The source code used to perform the experiments is publicly available\footnote{https://github.com/damicoedoardo/NNMF}.

\subsection{Baselines}
\label{subsec:baselines}

We compare the stability of the NNMF algorithms with the stability of the corresponding standard MF algorithm.
Moreover, we compare accuracy with some additional collaborative filtering baselines:

\begin{description}
\item[\itemknn, \userknn] Traditional item and user-based nearest neighbors approaches with cosine similarity. \cite{handbook}
\item[\slim] Linear regression model for top-n recommendation tasks with Bayesian Personalized Ranking as optimization function. \cite{slim} 
\item[\puresvd] Basic Matrix Factorization model based on SVD decomposition \cite{puresvd}. Notice that contrarily to the other MF methods we consider in this paper, it has an exact mathematical solution that can not include the similarity matrices introduced by NNMF. %Moreover it is not affected by the instability issues we are analyzing, since it does not involve a gradient descent training procedure.
\end{description}

All baselines have been tuned on a validation set by using a Bayesian optimizer.

\subsection{Datasets}
\label{sec:datasets}

We carry out experiments employing a number of research datasets:
\begin{description}
    \item[LastFM (\lastfm)] Implicit interactions gathered from the music website Last.fm. In particular, user \textit{listened} artist relations expressed as listening counts. \cite{cantador2011second}
    \item[Movielens1M (\movielensom)] Explicit interactions gathered from the website MovieLens. In particular, user \textit{rated} movie relations. \cite{movielens}%, i.e. records: user, movie, rating. \cite{movielens}
    \item[BookCrossing (\bookcrossing)] Explicit interactions gathered from the online book club BookCrossing. In particular, user \textit{rated} book relations. \cite{Bookcrossing}%, i.e. records: user, book, rating. \cite{Bookcrossing}
    \item[Pinterest (\pinterest)] Implicit interactions gathered from the social network Pinterest. In particular, user \textit{pin-to-own-board} image relations. \cite{pinterest}%, i.e. records: user, image, 1. \cite{pinterest}
    %\item[Epinions] One to five ranging ratings from the users of the consumer review website Epinions. \cite{epinions}
    \item[CiteULike (\citeulike)]  Implicit interactions gathered from the online scientific community CiteULike. In particular, user \textit{saved-to-own-library} paper relations. \cite{citeulike}%, i.e. records: user, paper, 1. \cite{citeulike}
\end{description}
We employ datasets with densities of interactions that range from 0.04\% to 2.51\%, as we want to take into account the effect of a varying density of the datasets on both the accuracy and stability of the models. 
The statistical details of the datasets are described in Table \ref{table:dataset_stats}.

%We propose datasets of various sizes in terms of user, item and interactions. This provide, on one hand, evidence about the capability to scale of the methods we propose, and, on the other hand, an analysis of the results with very diverse inputs. 

\begin{table}[t]\normalsize
\centering
\caption{Statistics of the evaluation datasets}
\small
\label{table:dataset statistics} 
\begin{tabular}{c|c|c|c|c}
\toprule 
\textbf{Dataset} & \textbf{Users} & \textbf{Items} & \textbf{Interactions} & \textbf{Density} \\
\midrule
\lastfm & 1859 & 2823 & 42798 & 0.81\% \\
\movielensom & 6038 & 3307 & 501114 & 2.51\%\\
% Mov20M & 138271 & 16905 & 9880969 & 0.42\% \\
\bookcrossing & 13975 & 33925 & 314499 & 0.06\%\\
\pinterest & 55186 & 9637 & 877796 & 0.16\%\\ 
%\epinions & 21554 & 25133 & 238353 & 0.04\%\\
\citeulike & 5536 & 15429 & 119919 & 0.14\%\\
\bottomrule 
\end{tabular} 
\label{table:dataset_stats}
\end{table}

The datasets used for the experiments have been preprocessed in two steps.
First, we brought all the interactions to either zero or one by means of thresholding, since BPR requires binary preference values. Among the implicit datasets, only \lastfm needs thresholding: we use threshold value equal to one, hence we convert every interaction to one if the user has listened at least once to an artist.
In explicit datasets we use threshold value equal to six for one to ten ratings and equal to three for one to five ratings.
Second we applied a filtering procedure keeping only users and items with at least five interactions, in order to remove entities with a too scarce amount of information.

\begin{comment}
\setlength{\tabcolsep}{2pt}
\begin{table}[H]
        \centering
             \caption{Parameters used for the preprocessing methods. For each dataset is reported the threshold selected for the impliciitization and the k used for the K-core procedure.} 

\begin{tabular}{ccccccc}
\toprule 
\textbf{Method} & \textbf{LastFM}&\textbf{Mov1M}&\textbf{Mov20M}&\textbf{BookCr} &\textbf{Pin}&\textbf{CiteUL}\\\midrule 
Thr & 1& 3& 3& 6& 1& 1 \\  \midrule
K-Core  & 5& 5& 5& 5& 5& 5\\ 
\bottomrule 
 \end{tabular}
 \label{table:dataset_prep}
\end{table}
\setlength{\tabcolsep}{4pt}
\end{comment}
Each dataset has been randomly partitioned performing a standard holdout procedure in three sets: train, validation and test accounting for 60, 20 and 20 percent of the available interactions.
%This type of split preserves the number of users and items present in each of the partition, necessary condition for the MFs algorithms, as they are able to provide recommendations only for the users and items present in the URM used for the training phase.

\subsection{Stability of representations}
\label{subsec:stability_representations}
\setlength{\tabcolsep}{2.4pt}
\begin{table}[t]
\centering     
\caption{Stability of representations @10 expressed as Jaccard index. Underline indicates the most stable algorithm. Bold indicates which is more stable between MF and NNMF.}
\small
\label{table:stability_representations}           
\begin{tabular}{c|cc|cc|cc|cc|cc}                 \toprule                 
\multirow{2}{*}{\textbf{Algorithm}} & \multicolumn{2}{c|}{\textbf{\lastfm}} & \multicolumn{2}{c|}{\textbf{\movielensom}} & \multicolumn{2}{c|}{\textbf{\bookcrossing}} & \multicolumn{2}{c|}{\textbf{\pinterest}} & \multicolumn{2}{c}{\textbf{\citeulike}} \\
%   & \multicolumn{1}{c}{Items} & \multicolumn{1}{c|}{Users} & \multicolumn{1}{c}{Items} & \multicolumn{1}{c|}{Users} & \multicolumn{1}{c}{Items} & \multicolumn{1}{c|}{Users} & \multicolumn{1}{c}{Items} & \multicolumn{1}{c|}{Users} & \multicolumn{1}{c}{Items} & \multicolumn{1}{c}{Users} \\ \midrule
  & \multicolumn{1}{c}{Item} & \multicolumn{1}{c|}{User} & \multicolumn{1}{c}{Item} & \multicolumn{1}{c|}{User} & \multicolumn{1}{c}{Item} & \multicolumn{1}{c|}{User} & \multicolumn{1}{c}{Item} & \multicolumn{1}{c|}{User} & \multicolumn{1}{c}{Item} & \multicolumn{1}{c}{User} \\ \midrule
 \bprmf & 0.73 & 0.67 & 0.70 & 0.70 & 0.47 & 0.32 & 0.45 & 0.30 & 0.57 & 0.58 \\
 \bprnnmf & \underline{\textbf{0.82}} & \underline{\textbf{0.87}} & \textbf{0.93} & \textbf{0.91} & \textbf{0.55} & \textbf{0.57} & \textbf{0.66} & \textbf{0.51} & \textbf{0.69} & \textbf{0.69} \\ \midrule
 \funkmf & 0.66 & 0.61 & 0.36 & 0.33 & 0.17 & 0.13 & 0.34 & 0.19 & 0.48 & 0.48 \\
 \funknnmf & \textbf{0.81} & \textbf{0.83} & \underline{\textbf{0.95}} & \underline{\textbf{0.92}} & \textbf{0.42} & \textbf{0.25} & \textbf{0.68} & \textbf{0.58} & \textbf{0.64} & \textbf{0.61} \\ \midrule
 \pmf & 0.62 & 0.56 & 0.55 & 0.47 & 0.38 & 0.26 & 0.32 & 0.19 & 0.60 & 0.40 \\
 \pnnmf & \textbf{0.76} & \textbf{0.78} & \textbf{0.81} & \textbf{0.68} & \underline{\textbf{0.57}} & \underline{\textbf{0.62}} & \underline{\textbf{0.63}} & \underline{\textbf{0.46}} & \underline{\textbf{0.75}} & \underline{\textbf{0.71}} \\ 
 \bottomrule \end{tabular} 
\end{table}
\setlength{\tabcolsep}{5pt}

Assessing the stability of representations requires to use a technique which is invariant to transformations on the vector space, such as permutation or rotation of coordinates. 
The representation of an item, or user, is stable if, in every new latent space, it is close to the same items, or users.
So we want to assess if the relationships among items and among users are maintained in the different embedding spaces and we can check this condition by ensuring that the neighborhoods formed in the new spaces are composed by the same set of users or items.
Note that the cross-entity relationships between users and items are checked with the stability of recommendations experiment in the next section.
%In fact, the generation of a recommendation list for a user is strictly related to the items which are closer to the user's one in the latent space.
%In this experiment, instead, we want to check the relationships among items or users themselves.

We compare MF and NNMF in a pairwise manner considering the differences in the stability of representations.
Every algorithm tested is executed ten times, changing the latent factor initialization, an effect obtained simply using different seeds for the random number generator used to assign the initial values. 
The order in which the training data samples are explored during the different runs of the algorithms, instead, is ensured to be constant.
We consider the latent factors representations learned and we create a list of the closest items to every item and users to every user, by means of a cosine similarity on the latent factor space. 
We compute these lists of $K$ nearest neighbors for the first of the ten models, both for MF and NNMF approaches. 
Then, we measure the degree of similarity of the $K$ nearest neighbors of the other nine models against the first considering the Jaccard index, a common statistic used to asses the similarity between sets.
%Given two sets $\mathcal{A}$ and $\mathcal{B}$, the Jaccard index $J(\mathcal{A}, \mathcal{B})$ between them is defined as:
%\[
%J(\mathcal{A}, \mathcal{B}) = \frac{|\mathcal{A} \bigcap \mathcal{B}|}{|\mathcal{A} \bigcup \mathcal{B}|}
%\]
%\begin{description}
%    \item[Jaccard index:] measures the percentage of overlapping recommendations across different runs.
%    \item[RBO index:] \cite{rbo} states how much the ranking of the items in the recommendation lists is preserved.
%\end{description}
Higher similarity of the nearest neighbors in the latent space across different runs suggests a higher similarity of the latent factor representations.
We performed the experiments using both $K=10$ and $K=100$ obtaining the same results, so, for brevity, in Table \ref{table:stability_representations} we report only the results with $K=10$.

The most evident trend is that we have a large improvement in the stability of the recommendations when applying the new NNMF method with respect to original MF algorithm in every configuration.
NNMF stability is over 50\% in almost all the experiments, and well above 80\% in two out of five datasets.
MF approaches, on the contrary, struggle to reach 50\% of stability for both users and items in three datasets out of five.
Analyzing the results, we observe higher stability for denser datasets, while it drops when the density of interactions is really low. 
%Indeed lowest values of stability on both users and items latent factors, are recorded in the three least dense datasets, namely \bookcrossing, \citeulike and \pinterest, while highest values are obtained in the two most dense datasets, \movielensom{} and \lastfm{}.
Among the MF algorithms, it is interesting to notice that the \bprmf approach (and its NNMF variant) have a better stability than the other methods on almost all the experiments.

%Another difference raises between the stability of users and items.
%MF algorithms tend to have more stable latent representations of items with respect to users' ones in almost every comparison.
%However, this trend is not reproduced with NNMF versions, where there is a better balance between the stabilities of items and users.
%Follows that the stability improvements for users representations are usually larger than the items' counterparts.

\begin{figure*}[t]
    \centering
    \begin{subfigure}[b]{0.27\linewidth}
        \caption{\bprmf vs \bprnnmf}
        \includegraphics[width=0.98\textwidth, keepaspectratio]{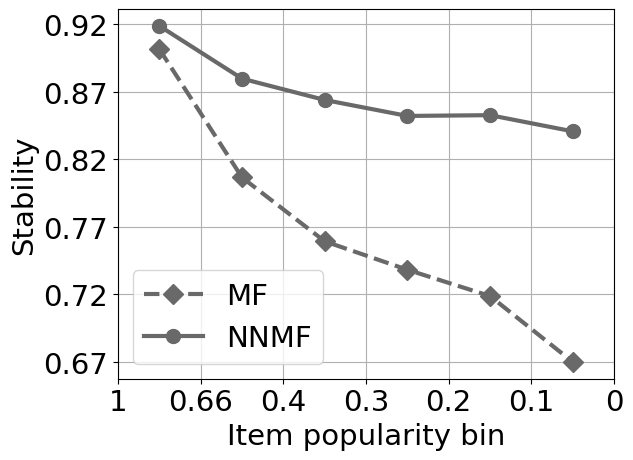}
    \end{subfigure}
    \begin{subfigure}[b]{0.27\linewidth}
        \caption{\funkmf vs \funknnmf}
        \includegraphics[width=0.98\textwidth, keepaspectratio]{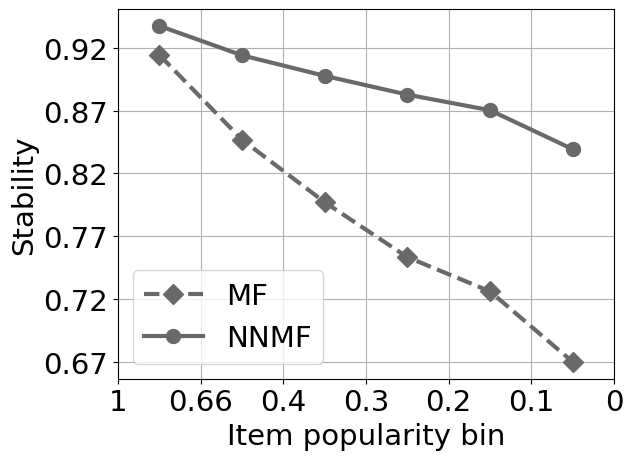}
    \end{subfigure}
    \begin{subfigure}[b]{0.27\linewidth}
        \caption{\pmf vs \pnnmf}
        \includegraphics[width=0.98\textwidth, keepaspectratio]{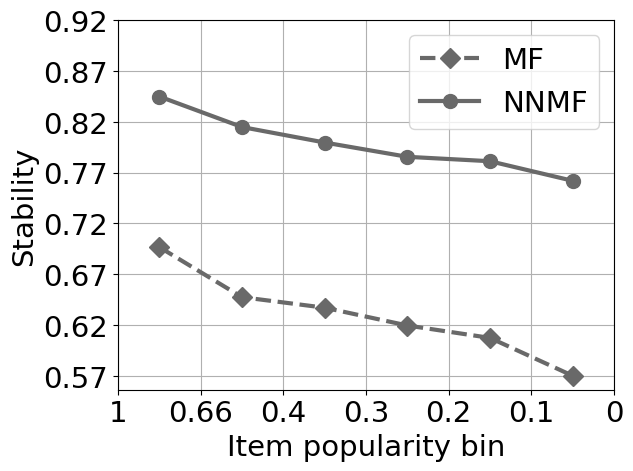}
    \end{subfigure}
     \caption{Stability of items representations for popularity ranges on the \lastfm dataset expressed as Jaccard @10. Every point represents the average stability value of the items in a bin. A bin contains all the items that belong to a range of popularity.}
    \label{fig:item_representation_stability}
\end{figure*}

As second step of this experiment, we also want to understand if a correlation between the popularity of an item and its stability exists, so we analyze the items' stabilities grouped by their popularity. 
%We define as long-tail all the least popular items that account for the 66\% of the interactions in the dataset.
We divide the items in 6 bins, depending on their popularity, using 1, 0.66, 0.4, 0.3, 0.2, 0.1 and 0 as thresholds.
%For instance, the leftmost bin reports the results for the 33\% most popular items, while the rightmost bin reports the result for the 10\% less popular items.
The first bin contains all the most popular items that account for the $1 - 0.66 = 34\%$ of the interactions in the dataset, i.e. the short-head\footnote{The short-head is defined as complementary to the long-tail, which is the set of less popular items that account for the 66\% of the interactions}.
The second bin contains all the most popular items, excluding those in the first bin, that account for the $0.66 - 0.4 = 26\%$ of interactions. And so on.
For each bin, we average the stability of the items that belong to it.

For brevity, in Figure \ref{fig:item_representation_stability} we show the results of the experiment performed only on the \lastfm dataset, but we obtained very similar results on all the datasets.
As expected, the stability of NNMF models is globally higher than the MF counterparts.
However, the plots show another clear trend: the representations of popular items are much more stable than unpopular ones.
This behavior is not surprising, since higher popularity determines a higher number of updates and, consequently, a more detailed representation supported by a higher amount of information. 
Instead, niche items representations are subject of few updates during the learning process, resulting in more fuzzy representations even at model convergence, where the impact of the initialization values used is still strong.
It is also evident that MF is subject to higher drops of stability, compared to NNMF, when passing from popular to niche items, widening the difference between them.
The proposed analysis proves that the recommendations made by MF models are noisy and strongly impacted the random initialization of the latent factors, since the available information about the real user's taste is often poor and marginally exploited.
NNMF models, instead, are overall more stable, and the stability is high also when considering non-popular items.
This has an impact also on the stability of recommendations and on the overall accuracy of the models, as we show in the following Sections.
%The propagation of updates to the neighborhood allows non-popular items to have a representation that results from a higher number of updates with respect to MF. Consequently, the representations become less dependent on the random initialization, and more aimed to represent user taste. Moreover, the latent representation of each user or item is not independent anymore, which means that also what we learn for a user's or item's neighbors contribute to its representation. Overall we obtain a higher stability and aimed recommendations of both popular and non-popular items.

\subsection{Stability of recommendations}
\label{subsec:stability_recommendations}
\begin{table}[t]
\small
\centering
\caption{Stability of recommendations @10 expressed as Jaccard index. Underline indicates the most stable algorithm. Bold indicates which is more stable between MF and NNMF.}
\begin{tabular}{c|c|c|c|c|c}                 
\toprule                 
\textbf{Algorithm} & \multicolumn{1}{c|}{\textbf{\lastfm}} & \multicolumn{1}{c|}{\textbf{\movielensom}} & \multicolumn{1}{c|}{\textbf{\bookcrossing}} & \multicolumn{1}{c|}{\textbf{\pinterest}} & \multicolumn{1}{c}{\textbf{\citeulike}} \\ \midrule 
 \bprmf & 0.70 & 0.78 & 0.28 & 0.50 & 0.49 \\
 \bprnnmf & \textbf{0.86} & \underline{\textbf{0.95}} & \textbf{0.52} & \textbf{0.69} & \textbf{0.65} \\ \midrule
 \funkmf & 0.72 & 0.60 & 0.05 & 0.28 & 0.31 \\
 \funknnmf & \underline{\textbf{0.87}} & \underline{\textbf{0.95}} & \textbf{0.18} & \textbf{0.75} & \textbf{0.51} \\ \midrule
 \pmf & 0.62 & 0.60 & 0.23 & 0.39 & 0.37 \\
 \pnnmf & \textbf{0.79} & \textbf{0.76} & \underline{\textbf{0.61}} & \underline{\textbf{0.78}} & \underline{\textbf{0.66}} \\ 
 \bottomrule \end{tabular} 
 \label{table:stability_recommendations}
\end{table}
As second experiment, we compare MF and NNMF in order to assess the differences in the stability of recommendations.
The experimental procedure is similar to the one described in last Section: every algorithm tested is executed ten times, changing the latent factor initialization.
But in this case we consider the top-10 recommendations provided by the ten models, both for MF and NNMF approaches, measuring the degree of similarity of the recommendations of the first model against the other nine, resorting again to the Jaccard index.
Table \ref{table:stability_recommendations} shows compatible results in the stability of recommendations with respect to what we observed for the representations in Table \ref{table:stability_representations}.
The recommendation lists generated by NNMF models are always widely more stable than their MF counterparts.
The MF implementations fail to reach the 50\% threshold for the Jaccard index in many configurations, with a negative spike of 5\% of \funkmf on \bookcrossing.
%This means that, considering recommendation lists composed by 10 items, in that configuration half of the users obtain completely different recommendations.
The NNMF versions largely mitigate this issue, as they are always able to outperform the respective MF implementations with a wide margin.
As for the representations, also in this case the difference between dense datasets and sparse ones is evident, with a higher stability in the first scenario.
%For example, in \movielensom, that is the most dense dataset, the algorithms have stability scores which are about three times those obtained on \bookcrossing, which is the least dense dataset.
Moreover, notice how the stability in both experiments tends to be higher for small datasets. 
This is not only related to the density of the dataset, since the number of items and users influence it, but having fewer items to take into account also reduces the different available options in the recommendation phase.
As final observation, among the MF algorithms note that the \bprmf approach is steadily better than the others on all the datasets, while among the NNMF this trend is not evident anymore.

\subsection{Accuracy}
\label{subsec:long_tail_accuracy}

\setlength{\tabcolsep}{2.2pt}
\begin{table*}[t]
\small
\centering
\caption{Long-tail accuracy. Bold indicates the best between MF and NNMF. Underline indicates the best performing algorithm.}
\begin{tabular}{c|cc|cc|cc|cc|cc|cc|cc|cc|cc|cc}
\toprule 
\multirow{3}{*}{\textbf{Algorithm}} & \multicolumn{4}{c|}{\textbf{\lastfm}} & \multicolumn{4}{c|}{\textbf{\movielensom}} & \multicolumn{4}{c|}{\textbf{\bookcrossing}} & \multicolumn{4}{c|}{\textbf{\pinterest}} & \multicolumn{4}{c}{\textbf{\citeulike}} \\
 & \multicolumn{2}{c|}{\textbf{MAP}} & \multicolumn{2}{c|}{\textbf{Recall}}& \multicolumn{2}{c|}{\textbf{MAP}} & \multicolumn{2}{c|}{\textbf{Recall}}& \multicolumn{2}{c|}{\textbf{MAP}} & \multicolumn{2}{c|}{\textbf{Recall}}& \multicolumn{2}{c|}{\textbf{MAP}} & \multicolumn{2}{c|}{\textbf{Recall}}& \multicolumn{2}{c|}{\textbf{MAP}} & \multicolumn{2}{c}{\textbf{Recall}} \\ 

 & @5 & @10 & @5 & @10& @5 & @10 & @5 & @10& @5 & @10 & @5 & @10& @5 & @10 & @5 & @10& @5 & @10 & @5 & @10 \\ 
 \midrule
 ItemKNNCF & 0.028 & 0.030 & 0.040 & 0.073 & 0.005 & 0.005 & 0.006 & 0.013 & \underline{0.014} & \underline{0.014} & \underline{0.018} & \underline{0.027} & \underline{0.013} & \underline{0.016} & \underline{0.023} & 0.044 & 0.047 & 0.051 & 0.065 & 0.110 \\ 
UserKNNCF & 0.020 & 0.022 & 0.032 & 0.065 & 0.005 & 0.006 & 0.007 & 0.022 & 0.007 & 0.007 & 0.010 & 0.016 & 0.010 & 0.013 & 0.019 & 0.037 & 0.047 & 0.052 & 0.071 & 0.116 \\ 
SLIM BPR & 0.024 & 0.026 & 0.038 & 0.068 & 0.002 & 0.003 & 0.003 & 0.010 & 0.004 & 0.005 & 0.006 & 0.010 & 0.010 & 0.012 & 0.017 & 0.035 & 0.045 & 0.050 & 0.071 & 0.114 \\ 
PureSVD & 0.027 & 0.028 & 0.045 & 0.084 & 0.022 & 0.021 & 0.020 & 0.047 & 0.002 & 0.002 & 0.003 & 0.004 & 0.005 & 0.007 & 0.010 & 0.022 & 0.018 & 0.019 & 0.022 & 0.046 \\ \midrule 
BPRMF & 0.035 & 0.037 & 0.050 & 0.087 & \textbf{0.022} & \textbf{0.022} & \underline{\textbf{0.025}} & \underline{\textbf{0.051}} & 0.006 & 0.007 & 0.009 & 0.014 & 0.010 & 0.012 & 0.018 & 0.035 & 0.037 & 0.042 & \textbf{0.060} & \textbf{0.099} \\
BPR NNMF & \underline{\textbf{0.039}} & \underline{\textbf{0.040}} & \underline{\textbf{0.058}} & \underline{\textbf{0.099}} & 0.017 & 0.016 & 0.018 & 0.039 & \textbf{0.009} & \textbf{0.008} & \textbf{0.011} & \textbf{0.016} & \textbf{0.012} & \textbf{0.015} & \underline{\textbf{0.023}} & \underline{\textbf{0.046}} & \textbf{0.039} & \textbf{0.043} & \textbf{0.060} & \textbf{0.099} \\ \midrule 
FunkSVD & \textbf{0.033} & 0.035 & 0.050 & 0.085 & 0.017 & 0.015 & 0.018 & 0.034 & 0.004 & 0.004 & 0.006 & 0.009 & 0.010 & 0.013 & 0.019 & 0.037 & 0.044 & 0.049 & 0.069 & 0.108 \\
Funk NNMF & \textbf{0.033} & \textbf{0.036} & \textbf{0.052} & \textbf{0.094} & \underline{\textbf{0.034}} & \underline{\textbf{0.028}} & \textbf{0.024} & \textbf{0.046} & \textbf{0.006} & \textbf{0.007} & \textbf{0.010} & \textbf{0.016} & \textbf{0.011} & \textbf{0.014} & \textbf{0.021} & \textbf{0.041} & \underline{\textbf{0.053}} & \underline{\textbf{0.058}} & \underline{\textbf{0.080}} & \underline{\textbf{0.127}} \\ \midrule 
ProbMF & 0.019 & 0.020 & 0.030 & 0.058 & 0.008 & 0.009 & 0.010 & 0.027 & 0.001 & 0.001 & 0.002 & 0.003 & 0.009 & 0.011 & 0.016 & 0.033 & 0.033 & 0.038 & 0.052 & 0.090 \\ 
Prob NNMF & \textbf{0.034} & \textbf{0.036} & \textbf{0.054} & \textbf{0.095} & \textbf{0.027} & \textbf{0.024} & \underline{\textbf{0.025}} & \textbf{0.048} & \textbf{0.007} & \textbf{0.007} & \textbf{0.009} & \textbf{0.014} & \textbf{0.011} & \textbf{0.013} & \textbf{0.020} & \textbf{0.038} & \textbf{0.047} & \textbf{0.049} & \textbf{0.064} & \textbf{0.107} \\
 \bottomrule
\end{tabular}
\label{table:top_n_5_MAP_0.66}
\end{table*}
\setlength{\tabcolsep}{5pt}

% \begin{table*}[h]
% \centering
% \caption{Long-tail accuracy expressed in terms of MAP@5. Bold indicates who is best among MF and NNMF pairwise, in case the difference is statistically significant. Underline indicates the best performing algorithm.}
% \begin{tabular}{c|cccccc}
% \toprule 
% \textbf{Algorithm} & \textbf{LastFM}&\textbf{Mov1M}&\textbf{BookCr} &\textbf{Pin} & \textbf{CiteUL}\\\midrule 
% ItemKNNCF & 0.0280 & 0.0047 & \underline{0.0138} & \underline{0.0134} & 0.0471 \\  
% UserKNNCF & 0.0197 & 0.0051 & 0.0067 & 0.0093 & 0.0465 \\ 
% SLIM BPR & 0.0243 & 0.0018 & 0.0043 & 0.0096 & 0.0449 \\  
% PureSVD & 0.0267 & 0.0218 & 0.0019 & 0.0055 & 0.0185 \\ \midrule 
% BPRMF & 0.0340 & \textbf{0.0259} & 0.0059 & 0.0093 & 0.0294 \\ 
% BPR NNMF & 0.0356 & 0.0179 & \textbf{0.0071} & \textbf{0.0118} & \textbf{0.0306} \\ \midrule 
% FunkSVD & \textbf{\underline{0.0366}} & 0.0116 & 0.0031 & 0.0109 & 0.0437 \\ 
% Funk NNMF & 0.0340 & \textbf{0.0325} & \textbf{0.0050} & \textbf{0.0114} & \textbf{\underline{0.0479}} \\ \midrule 
% ProbMF & 0.0146 & 0.0094 & 0.0023 & 0.0078 & 0.0251 \\ 
% Prob NNMF & \textbf{0.0339} & \textbf{\underline{0.0353}} & \textbf{0.0054} & \textbf{0.0091} & \textbf{0.0362} \\ 
% \bottomrule \end{tabular}
% \end{table*}

%\subsubsection{Results}
%\label{subsubsect:results_long_tail}

%\subsubsection{Experiment description}
%Recommending popular items is trivial and does not bring much benefits to users and content providers. 
%On the other hand, recommending less known items adds novelty and serendipity to the users but it is usually a more difficult task \cite{puresvd}.
Recommending non-popular items adds novelty and serendipity to the users, but it is usually a more difficult task compared to the recommendation of popular ones \cite{puresvd}.
In this experiment we measure the accuracy of MF and NNMF in suggesting non-trivial items with a standard long-tail accuracy experiment, \textit{i.e.} we compute the top-n performance of the algorithms using as ground truth the long-tail of the test set, while the training set is considered at its whole.
We define as long-tail all the least popular items that account for the 66\% of the interactions in the dataset.
We evaluate the behavior of NNMF and MF approaches in this scenario, carrying out pairwise comparisons.
Moreover, to provide context to our results, we additionally score other competitive collaborative baselines \cite{mauri} described in Section \ref{subsec:baselines}.
In Table \ref{table:top_n_5_MAP_0.66} we report the performance obtained by the different algorithms, expressed by the Mean Average Precision and the Recall at two cutoffs, 5 and 10.
In almost 90\% of the measures, the NNMF algorithms perform better than or equal to the corresponding MF versions, and in more than 80\% of the measures, the improvement provided by NNMF is consistent.

%Overall, the first interesting result is that MF techniques prove to be unable to outperform the baselines in a number of cases, especially in the most sparse datasets.
%For example, let's consider \bookcrossing, the dataset with lowest density (Table \ref{table:dataset_stats}): \itemknn, \userknn and \slim outperform MFs.
%In particular, \itemknn doubles the MAP of the best performing MF approach.
%When the datasets are really sparse, MF struggles, since the number of interactions on the long-tail is lower: the non-popular items are not modelled properly, and the recommendation quality falls, as expected.
%On the contrary, in the datasets with highest density, \movielensom and \lastfm, MF models are able to outperform most of the baselines.

%A surprising result is that the \itemknn model is able to obtain an outstanding accuracy even on the long tail, despite it is known to be a strongly popularity biased algorithms, and it is the best performing algorithm on two datasets.

\begin{figure}[t]
    \centering
    \begin{subfigure}[b]{0.49\linewidth}
        \caption{BPR-Opt}
        \includegraphics[width=0.98\textwidth, keepaspectratio]{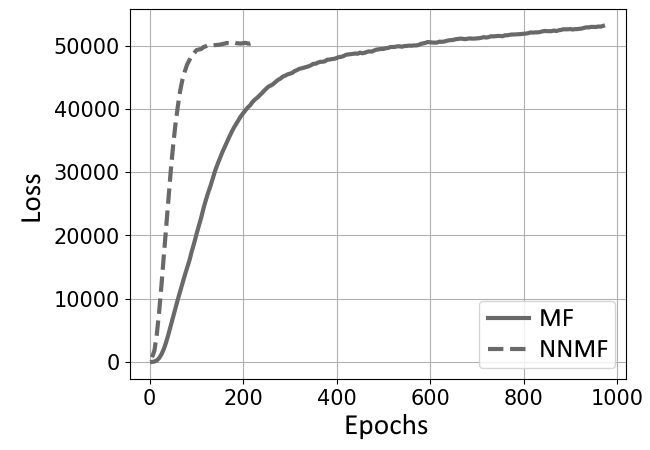}
    \end{subfigure}
    \begin{subfigure}[b]{0.49\linewidth}
        \caption{MAP}
        \includegraphics[width=0.98\textwidth, keepaspectratio]{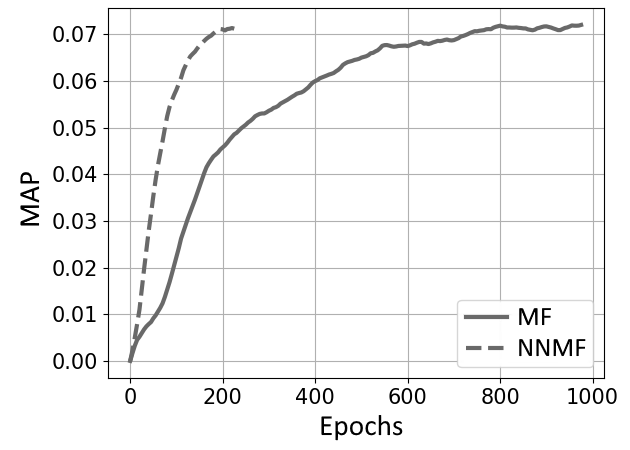}
    \end{subfigure}

    \caption{Comparison between the training procedures of \bprmf and \bprnnmf on the \citeulike dataset. The plot on the left represents the BPR-Opt value on the training set. The plot on the right represents the MAP@5 obtained on the validation set.}
    \label{fig:training_mf_nnmf}
\end{figure}

NNMF models achieve highest accuracy on three datasets over five and in the remaining cases they try to fill the gap between the best performing model (usually \itemknn) and classic MF algorithms, proving to be at least competitive against the other models across the datasets.
Indeed, notice that even when a NNMF model is not the most accurate model on the long tail, it is the second best performing algorithm.
The poor long-tail accuracy of MF obtained in many scenarios is quite surprising, especially if we consider that MF approaches are known for being less popularity biased than other CF approaches \cite{abdollahpouri2019unfairness, channamsetty2017recommender, whatrecommendersrecommend}.
We can conclude that these models are able to recommend niche items, but they are often noisy recommendations, resulting from the low number of updates on the latent factors of non-popular items, rather than a real evidence of the user's taste.
NNMF models, instead, leverage the knowledge of the neighborhood to construct more consistent item representations on the long-tail, transforming its part of non-popular recommendations in higher quality long-tail accuracy.\\
%A last mention to the accuracy on the whole test set (Appendix \ref{appendix:topn-performance}). It shows that NNMFs have accuracy which is comparable with the other methods, if we evaluate on long-tail and short-head together. This proves that NNMF methods manage to achieve a good modelling of both long-tail and short-head, and that they appropriately consider both parts during the recommendation phase, staying competitive with other recommendation approaches in the standard top-n recommendation task.
\subsection{Model training}
\begin{comment}
\setlength{\tabcolsep}{2.0pt}
\begin{table}[t] 
\centering 
\caption{Training time in minutes. We report average training times and standard deviations across ten runs with optimal hyperparameter configurations.}
\small
\begin{tabular}{c|c|c|c|c|c} \toprule \textbf{Algorithm} & \textbf{\lastfm} & \textbf{\movielensom} & \textbf{\bookcrossing} & \textbf{\pinterest} & \textbf{CUL} \\ \midrule 
\bprmf & 1.4 ± 0.01 & 8.2 ± 0.68 & 8.7 ± 0.89 & 28.2 ± 2.51  & 4.9 ± 0.60 \\
\bprnnmf & 5.1 ± 0.13 & 6.0 ± 0.02 & 29.2 ± 0.178 & 70.2 ± 2.04 & 2.4 ± 0.02 \\ \midrule 
\funkmf & 0.5 ± 0.01 & 0.4 ± 0.01 & 6.7 ± 0.08 & 20.9 ± 0.04 & 1.9 ± 0.01 \\
\funknnmf & 1.0 ± 0.01 & 100.3 ± 1.31 & 10.2 ± 0.05 & 42.7 ± 0.34 & 4.7 ± 0.02 \\ \midrule 
\pmf & 1.2 ± 0.01 & 6.4 ± 0.02 & 10.8 ± 0.06 & 26.5 ± 0.08 & 8.5 ± 0.05 \\ 
\pnnmf & 2.4 ± 0.01 & 6.5 ± 0.16 & 62.5 ± 0.48 & 40.4 ± 0.34  & 4.4 ± 0.02 \\ \bottomrule 
\end{tabular} \label{table:training_time} \end{table}
\setlength{\tabcolsep}{5pt}
\end{comment}

As last experiment, we also investigated the behavior of the models during the training procedure.
For brevity, in Figure \ref{fig:training_mf_nnmf} we show, as an example, the comparison between \bprmf and \bprnnmf on the \citeulike dataset, but we could observe the same trend in all the other configurations.
The plot on the left shows the maximum posterior estimator, called BPR-Opt in \cite{BPR}, on the training set, while the one on the right shows the performance on the validation set, expressed as the MAP@5.
Even if the starting and the convergence values of both the BPR-Opt and the MAP are quite similar between the two models, the NNMF version of the algorithm reaches lower BPR-Opt values and higher performance in a largely smaller amount of epochs.
Indeed, notice that the NNMF version reaches convergence after about 200 epochs and the training is interrupted by the early stopping technique, while the original MF version needs about 1000 epochs to reach the same performance.
%Even if the starting and the convergence values of the MAP are quite similar between the two models, the NNMF version of the algorithm reaches higher BPR-Opt values and higher performance in a largely smaller amount of epochs.
%Moreover, notice that the NNMF version reaches convergence after about 500 epochs and the training is interrupted by the early stopping technique, while the original MF version needs about 900 epochs to reach the same performance on the validation set, but a BPR-Opt value 20\% lower.
This result proves that the propagation of the information we have about a user or an item also to its neighbors is very effective and useful, allowing the model to reach the optimal performance in a lower number of iterations over the training data.

\section{Conclusions and Future Works}
\label{sec:conclusions}

%MF approaches are known to be less biased towards the recommendation of popular items, with respect to the majority of collaborative filtering approaches.
%However, we have shown that the long tail accuracy of these recommendations is not always as competitive as expected, even against simple popularity biased baselines like \itemknn, that, instead, perform surprisingly well even in this scenario.
%Moreover, we posed another issue of MF approaches related to the stability of recommendations generated and the representations of items and users in the latent space.
%We have demonstrated that by simply changing the initial values of the latent factors, very often more than half of the items in the recommendations lists generated by the same configuration of the same algorithm trained on the same data, change, and we have the same effect also in the neighborhoods of items and users.

In this paper we present \textit{Nearest Neighbors Matrix Factorization}, a generalization of classic Matrix Factorization.
The new framework merges nearest neighbors and Matrix Factorization techniques in order to mitigate the drawbacks induced by the scarcity of collaborative information available, especially for unpopular items.

The results of extensive experiments on five different datasets show that classic MF approaches are particularly affected by instability.
By simply changing the initial values assigned to the latent factors of users and items, the same model, trained on the same data and with the exactly same configuration, provides very different recommendations and latent representations of users and items at convergence, two issues that we call, respectively, \textit{instability of recommendations} and \textit{instability of representations}.
Moreover, we show that exists a correlation between the popularity of an item and the stability of its representation.

To assess the validity of the new technique, we propose the NNMF extensions of three of the most common MF algorithms, checking the accuracy and the stability of the new models.
The NNMF approaches provide large and consistent stability improvements in every scenario, and they are also able to increase the accuracy of their MF counterparts in almost every configuration, certifying the quality of the proposed framework.
Finally, we show that the new models are able to reach convergence in a fraction of the epochs necessary to the MF approaches, thanks to the propagation of the information through the neighborhood relations. 

Future works are addressed towards the study of the stability of more complex embedding-based models.
%Moreover, it is possible to explore different possibilities in the selection or the learning of the similarity matrices for users and items, to further improve the accuracy and the stability of the models.
%Indeed, the proposed implementation of NNMF uses a simple shrinked cosine similarity, which is fixed and does not explicitly take into account many aspects that can be crucial for the optimization of the performance, like the singularity of a user's taste (i.e. the degree of independence of items and users), or even different types of information, when available.

\section*{Acknowledgement}
Giovanni Gabbolini and Edoardo D'Amico would like to acknowledge a grant from Science Foundation Ireland (SFI) under Grant Number 12/RC/2289-P2, which is co-funded under the European Regional Development Fund, that partially supported this research.

\bibliographystyle{ACM-Reference-Format}
\bibliography{sample-base}

%\clearpage
%\appendix
%\input{appendix.tex}

\end{document}